# Refined Parameters of Chelyabinsk and Tunguska Meteoroids and their Explosion Modes

**Yury I. Lobanovsky**

*IRKUT Corporation*

*68, Leningradsky prospect, Moscow, 125315, Russia*

*E-mail: streamphlow@gmail.com*

**Abstract**

This paper describes application of mathematical model that establishes relationship between parameters of celestial bodies motion in the spheres of activity of the Sun and the Earth with mass-energy characteristics of these objects and their explosion modes during destruction in the Earth atmosphere, that in turn are linked with phenomena observed on underlying surface. This model was used to calculate the characteristics of objects that caused the Chelyabinsk and Tunguska explosions with using of its trajectory parameters described in recent scientific publications (late 2013 – early 2014). It turned out that the size of Chelyabinsk meteoroid was equal to 180 – 185 meters, and its mass was close to 1.8 megatons. Energy of its explosion was equal to 57 megatons of TNT, size of Tunguska meteoroid was equal to 105 m, mass – 0.35 megatons, while energy of explosion was about of 14.5 megatons of TNT. Due to the common origin of these two celestial bodies their average density was equal – about of 570 kg/m$^3$.

## I. Introduction

Within one and a half month after fall of Chelyabinsk meteoroid that occurred February 15, 2013, there was created a mathematical model that relates parameters of celestial bodies motion in spheres of activity of the Sun and the Earth with mass-energy characteristics of these objects and their explosion modes during the destruction in the Earth atmosphere, that in turn are linked with phenomena observed on the underlying surface [1, 2]. This model was used then for calculation of characteristics of Chelyabinsk and Tunguska meteoroids that in these papers, as well as in subsequent articles of the author (see below) were linked by their origin and, therefore, the proximity of their orbits. Thus, main data were obtained which fully describe these remarkable phenomena through regular physical and mathematical procedure without any speculative hypotheses and/or assumptions.

However, those calculations were based on preliminary information obtained during first 2 – 3 weeks after the Chelyabinsk incident, and author's own estimations of certain key parameters of this phenomenon, which, for obvious reasons, could not have a high accuracy at that time. In addition, over time, some internal parameters of the mathematical model, such as nominal height of the atmosphere, as well as its characteristic height, where air density is changed in e times (where e is the base of natural logarithm), in the framework of this problem have also been refined. After executing of much number of calculations there were used its values more closely related to the conditions of the problem and to beginning of the relevant processes of heating and ablation of material from surface of objects when they enter the Earth atmosphere. This parameter was calculated now iteratively, and for Chelyabinsk and Tunguska meteoroids it was approximately equal to 90 km (more details on this issue will also be discussed below). Thus, it was refined the influence of the Earth atmosphere on meteoroid trajectory in framework of numerical model at the cost of some complication of computational procedure.

In addition, the characteristic height in calculating the trajectory was the same (h = 8.00 km) as the corresponding parameter in «external» interactive module, which describes the destruction of meteoroids in the atmosphere [3] and was developed a few years earlier by other researchers (see [3, 4]). These, in principle, relatively small changes in the internal parameters of the model have some influence on the results of calculations. Moreover, due to the significant non-linearity of the model there may be that at a certain set of parameters the solution not exists under the old, rougher model, and exists in the model refined. This occurs at the boundaries of existence of a solution, which, like in any other non-linear problem can be found not always.

Two big review articles [5, 6] on Chelyabinsk incident were published in November 2013 in famous journals – Nature and Science by two large scientific groups (33 and 59 co-authors, respectively). These articles finally were given sufficiently reliable and relatively detailed data, derived from video images and eyewitness reports, about the parameters of meteoroid's trajectory and the phenomena which accompanied its passage and explosion. After this came the opportunity to recalculate the characteristics of Chelyabinsk and Tunguska meteoroids on these materials, refining the observed pattern of the phenomenon in the sky at Chelyabinsk in comparison with data from express reports in the media and fragmentary personal reports of eyewitness used in papers [1, 2]. The recalculation results of characteristics of Chelyabinsk and Tunguska meteoroids on these data, with the aid of the algorithm described in [1, 2] and with taking into account the changes mentioned in the introduction, are presented in this paper.

## II. Way to obtain the most accurate and precise trajectory parameters of Chelyabinsk meteoroid

It is known that the orbital parameters of Chelyabinsk meteoroid which have been received in the first 2 – 3 weeks after the incident and were the input data of the calculations, the results of which are presented in [1, 2], characterized by considerable scatter (see, for example, [7]). Data available now are more accurate, however, comparison of the results of several recent sources shows that the variation of parameters decreased, but remained quite noticeable (see [5, 6, 8, 9]). Therefore, when new numerical calculations were executed, main attention should be directed on the reliability and accuracy of the input parameters for the numerical algorithm. For this, after consultations with experts-astronomers there was chosen the speed of the object's entry into the atmosphere as one of three required input parameters describing the orbit of Chelyabinsk meteoroid before the collision with the Earth (not counting the known data on the intersection of orbits of the object and the Earth). This parameter was obtained directly from the videos and, therefore, in principle, contains minimal errors from algorithms for computing and converting. The second quantity – value of semi-major axis with a high degree of accuracy is obtained from resonance 13:6 with the Earth orbit (see [1, 2]), virtually indisputable evidence of which were presented in paper [10]. Thus, the value of semi-major axis of Chelyabinsk meteoroid a = 1.674 AU (astronomical unit) is known from its period of revolution of $\tau = 13/6 = 2.167$ years ($a \sim \tau^{2/3}$).

At zero angle of inclination of meteoroid's orbit plane to the ecliptic and with known point of its intersection with the Earth orbit of these two parameters are enough to determine the orbit of the object. As the angle of orbit inclination was little according to all data sources, its effect on input parameters in the atmosphere is small. Therefore, a little rounded value i = 5.0° was just taken from source [6].

Therefore, it remains to analyze the latest published data of Chelyabinsk meteoroid entry speed v in the Earth atmosphere. And here again there were some problems in connection with noticeable differences even these new and revised data. The smallest value of the speed and the smallest error were stated by G. Ionov – v = 18.85 ± 0.09 km/s for two series of measurements [8]. Were also given the following values for this parameter: v ≈ 19.0 km/s [5], v = 19.16 ± 0.15 km/s [6], and v = 19.3 ± 0.9 km/s [9]. Taking into account that the accuracy of the results of the source [9], in which has been described a new method of working out the trajectory measurements, as quite clearly states the author of this work, is currently lower than in other studies mentioned here, this value was excluded from this comparative review. Yet even from data sources [5, 6, 8] is followed that the accuracy of the determination of the object's entry speed into the atmosphere is not better than 0.3 km/s – v = 19.0 ± 0.3 km/s.

Such variations in the speed when errors mentioned in sources [6, 8] not exceed ± 0.15 km/s, suggests that at least one of these results has a systematic error not less than 0.15 km/s. Probably that it is associated with application to a little sloping and long trajectory of Chelyabinsk meteoroid standard algorithms for sufficiently steep and/or short trajectories («flat» Earth, the trajectory in the atmosphere is a straight line segment), which are fully adequate only for objects with the scale much smaller than the scale of Chelyabinsk meteoroid. Other noticed problems and inconsistencies in the data of paper [6], which are discussed in a separate article devoted exclusively to criticism of sources [11], allowed us to conclude that this error was there. It follows that the most accurate data of entry speed were obtained by G. Ionov, who has made a video of Chelyabinsk bolide flight almost from the doorstep of his house and then repeatedly made a photographs of night sky with the same position, having reduced the random errors of measurements to level of ± 0.09 km/s [8].

Thus, the input speed of Chelyabinsk meteoroid according to the source [8] in the numerical calculations was 18.85 km/s, that is on 1.33 km/s greater than in earlier calculations [1, 2] (the perihelion of the meteoroid's orbit fell from q = 0.80 ± 0.02 AU down to 0.746 AU, aphelion increased from Q = 2.55 up to 2.603 AU and eccentricity changed from e = 0.52 to 0.554 AU, while maintaining value of the semimajor axis). It may be noted that the value of semi-major axis of Chelyabinsk meteoroid used in this paper differs from the average value based on data from four papers given in source [6] (a = 1.70 ± 0.05 AU) on – 1.6 %, and from the result of source [9] (a = 1.67 ± 0.10 AU) – on + 0.2 %. Values of perihelion are follows: average value from source [6] – q = 0.77 ± 0.05, that is, the difference amounts + 3.2 %, from source [9] – q = 0.73 ± 0.01, and the difference amounts – 2.1 %. So, there is a perfect agreement between all of these data.

In addition, the basic version of computational results with v = 18.85 km/s was recalculated to a velocity equal to 19.00 km/s, and the consequences of this possible increase of speed were analyzed in the article.

## III. Way to obtain the most reliable and accurate data describing the Chelyabinsk explosion

Now it is necessary to clarify the quantitative characteristics of the phenomena associated with the meteoroid's approach to «point» of explosion, and with the explosion itself. The explosion of Chelyabinsk object was occurred because of its destruction down to small crumbs and dust, and very sharp braking of debris avalanche. From observations follows that all these process did not occur instantly. Two peaks of electromagnetic radiation were observed during this explosive process. From data of source [6] follows that the second main peak of emission was recorded at an altitude of about 30 km, and the completion of the fireball formation before its conversion to relatively little luminous cloud occurred at the altitude of approximately 27 km. From the description of nuclear explosions is known that the shock wave separation from fireball occurs in that moment (see, for example [12]), and the shock wave of meteoroid broke away from the fireball at this height range in the vicinity of its lower border.

The mathematical model of the explosion used in the described computational method is simplified. The explosion of a celestial body is similar to nuclear burst in this model that is this explosion is spherically symmetric and happens instantly. Therefore, we should specify «point» of explosion, which allows the best way to approximate much more complicated and lengthy process of explosion to this simplified model of burst. This may be done only by varying the main parameters of the explosion point in a reasonable range of variation, in process of comparing the calculated and the observed parameters characterizing the propagation of a shock wave, to give the best agreement between the calculated and experimental data. We may assume that the height of the explosion point lies in the range 27 – 30 km.

The average value of the geographical coordinates for the trajectory of Chelyabinsk meteoroid corresponding to the middle of this very short section of the path is very close to the data from sources [1, 2, 6, 9]. North latitude is equal to 54.87° and east longitude is 61.20°. Deviation in latitude from the data used in papers [1, 2] is 0.02°, and for longitude differences there were no quite. Estimated time of the explosion has not changed – 9:20:30 February 15, 2013. Estimates of the geodesic trajectory azimuth were 273.2° [6] – this means that the object was moving from east to west, shifting northward at 13.2°, which is 1.3° less than in earlier calculations, with the azimuth determined according to the meteoroid's trace [1, 2].

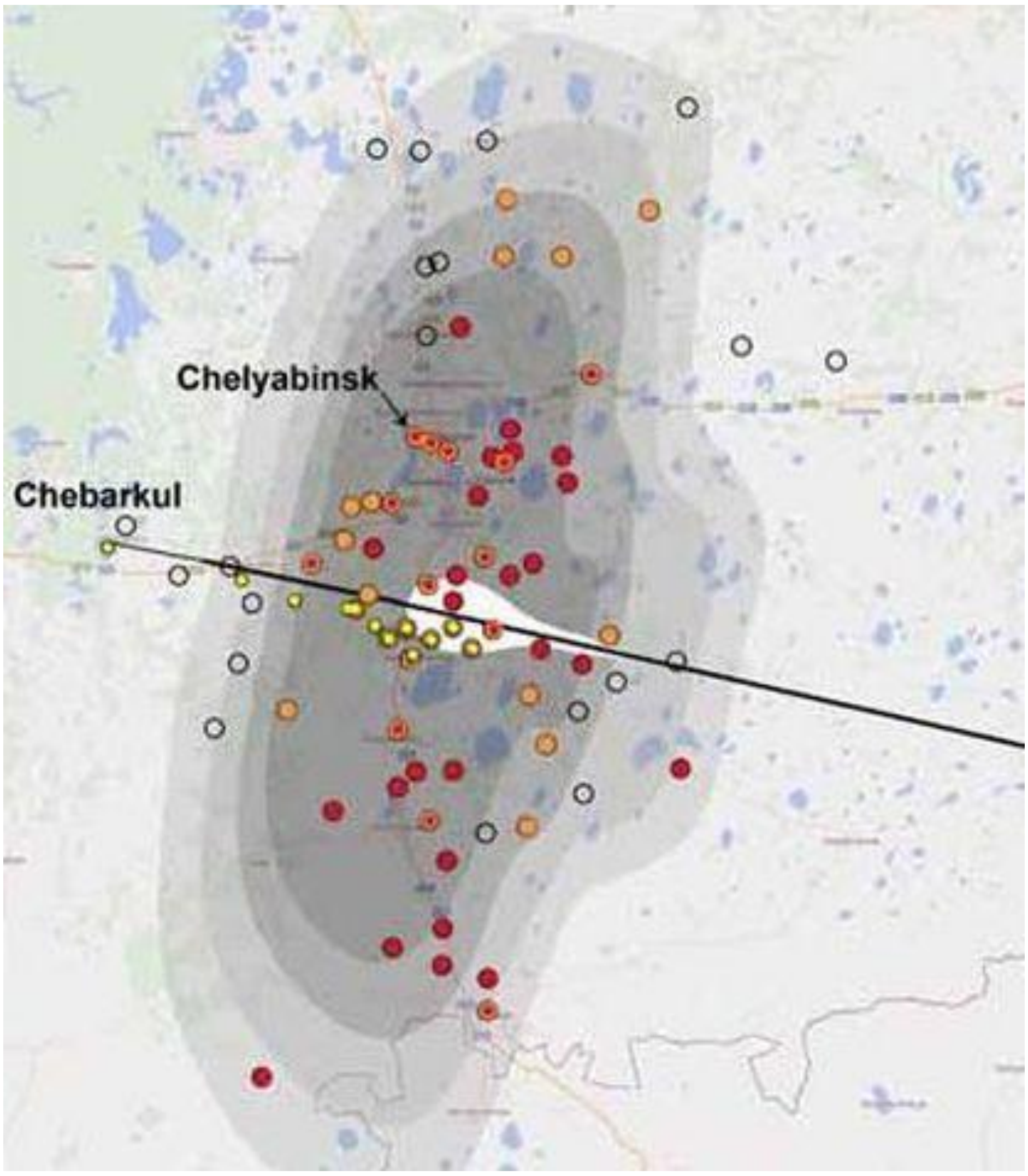

Fig. 1

In order to start the numerical calculations in the framework of this model, we should to determine one more parameter – anywhere but far enough from the epicenter of the explosion is necessary to know the overpressure peak on a shock wave. As in earlier papers by the author [1, 2], so in articles [5, 6] published much later, parameters of such type were determined through a state of glass windows in the zone of destruction, that is, the presence and/or part of shattered windows in the area of the shock wave action. Such map is given in reference [6] and its main part is shown in Fig. 1. Red points (according to the Emergency Department data), as well as orange points (from field surveys of source [6] co-authors) are shown for regions of Chelyabinsk and for localities of the homonymous province, where the window glasses were shattered. The open points mark locations where noticeable quantities of shattered glass were not registered. Yellow points on this map show drops of small meteoroid's fragments, and there is no interest to them for us in the present context.

It can be seen that the zone with shattered windows is something like a few rounded rectangle. Black line is a projection of the meteoroid trajectory on the Earth surface. The distances on interactive Yandex-map from the epicenter to the maximum distant settlement with shattered windows in the direction perpendicular to the projection of the flight path was 105 km. That is the maximum size of the zone of broken glass in this direction was about 210 km, and along the trajectory its size was 2.2 times less – about 95 km.

The white area in Fig. 1 illustrates the change in the emission intensity of the fireball on the flight path that is light curve of bolide [6]. Gray areas varying intensities show borders of zones with constant values of overpressure peak on the shock wave, obtained by calculations of explosion by using the numerical gas-dynamic code [13], provided that the energy of the explosion was distributed along the flight path proportionally to the light curve [6]. There is if not quantitative, then at least a qualitative similarity between these boundaries and the limit of zone with broken glass. At explosion in a point, despite the small slope of meteoroid trajectory, this gasdynamical code as well as other similar methods gives almost circular picture of boundaries with constant values of overpressure peak on the shock wave (see [1, 2, 6]).

Data presented in [6] suggest that here, unlike of Tunguska explosion, in which explosion height was in more than three times lower (see [1, 2]), the impact of ballistic shock wave on the underlying surface was almost indistinguishable. Further, from the fact that the slope of the trajectory of Tunguska meteoroid was in several times larger than for Chelyabinsk object, we may conclude that the distance of the explosive energy release in the first case have to be much shorter and, therefore, Tunguska burst was considerably closer to the explosion in the point. Therefore, there is every ground to conclude that the reasons for deviations from circular symmetry of destruction zones at Tunguska and Chelyabinsk incidents are different, and to understand why they are so different in their forms. From this also follows that the boundary condition for Tunguska explosion – overpressure peak on the shock wave for tree felling is equal to 30 kPa at the distance of 20 km from the epicenter [1, 2], does not need to change.

And here, at Chelyabinsk, the approximation of real border of destruction zone with the aid of circular symmetrical region, required for the algorithm used in the calculation module [3], leads to the following boundary condition – the overpressure peak on the shock wave, which is required for certain multiple glass broken, equal 5.0 kPa [14], is achieved at a distance of 80 km from the epicenter, what is on 10 km less than in the original calculations (see [1, 2]). This distance was obtained by equating of areas of real and circular symmetric approximation of destruction zones, what is a fairly obvious way to do such computational assessment. Test calculations then were carried out at values of the radius of this zone from 74 to 86 km.

But another source gives much more detailed information about the overpressure peaks on the shock wave at the points with known exact coordinates. In reference [5] was reported that part of broken glass in the area around the Chelyabinsk Zinc Plant is indicated on «overpressures close to 7 – 8 kPa». Location of explosion epicenter on the map of Chelyabinsk and its environs (mark 1) and Plant's storehouse of zinc concentrate (mark 2), where there was destruction of the roof, is shown in Fig. 2. The distance between them is 39.5 km. It should be noted that the Ice Palace «Urals Lightning» (mark 3), in which in the morning February 15, 2013 one supporting beam has collapsed, several beams were curved, and cladding from the facade was destroyed (not to mention the broken glass) [15], was located at a distance of 35 km from the epicenter, and almost on the same line, which connects the epicenter of the explosion with a storehouse of zinc concentrate (deviation from this line is not more than 0.65 km, see Fig. 2).

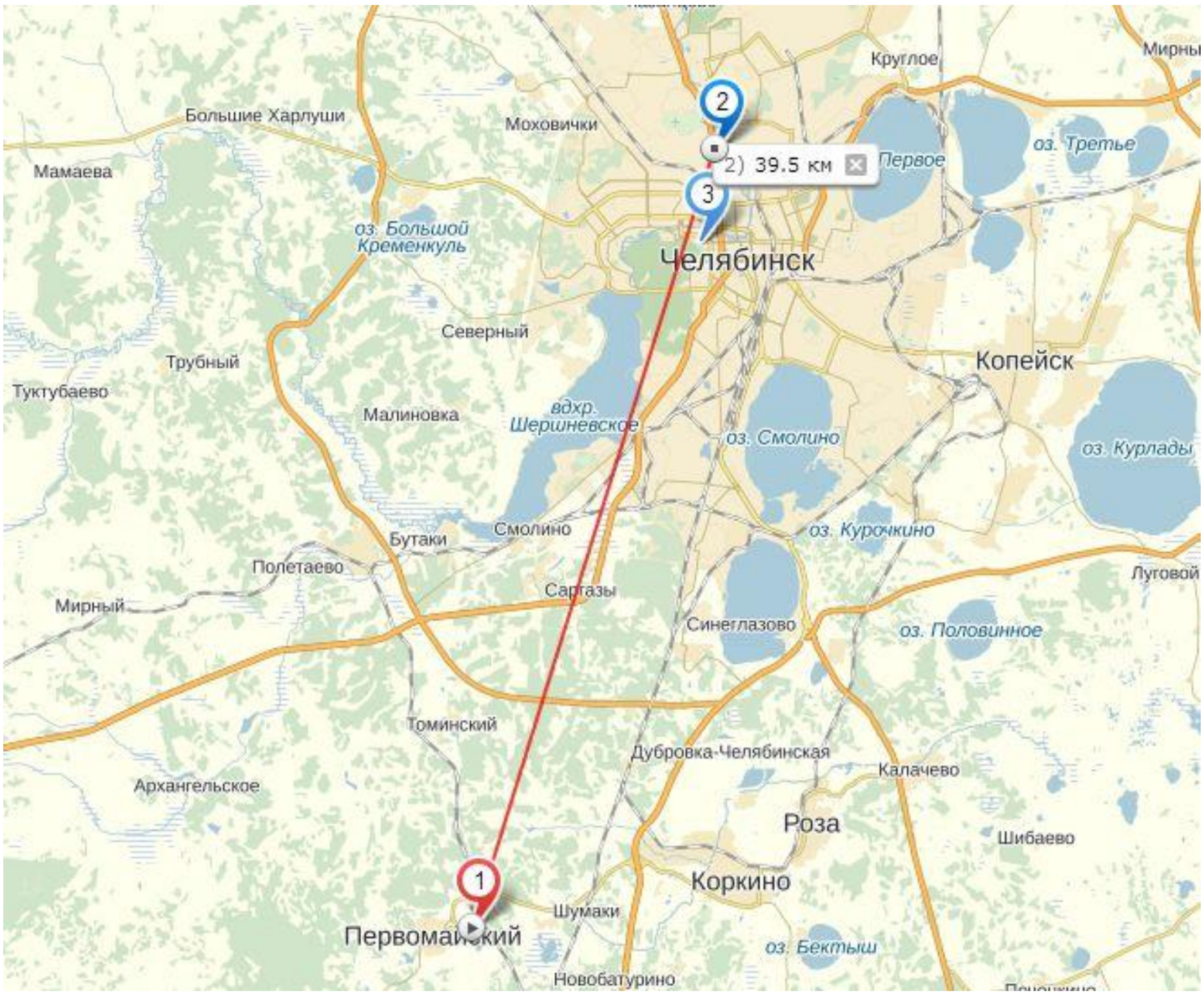

Fig. 2

The level of overpressure on the shock wave «in the Chelyabinsk urban area based on all forms of window damage» was estimated in source [5] as 3.2 ± 0.6 kPa, and the overpressures of 7.5 ± 0.5 kPa in the region, which is extending from the epicenter of the explosion on a few kilometers further than center of the city, were surprise for authors of this paper. They tried to explain this fact using such terms as «caustic» and «constructive acoustic interference». However, words, no matter how are profound they would be on their own, without disclosing the real mechanisms of the phenomenon cannot explain anything. For a person who is familiar with reflection and interference of nonlinear shock waves, which are qualitatively different from reflection and interference of linear acoustic waves, these mechanisms are transparent enough. But, there was none such person of 33 co-authors of the article from reference [5], apparently. In this context, it is worth noting that first article of the author of this work was devoted to the three dimensional interference of shock waves [16].

Elementary acquaintance with features of propagation of shock waves leads to a quite obvious thought that a shock wave from the airburst, height of which is comparable and even greater than the distance to the target, interacts with it not so, as a shock wave from low-altitude explosion, when the removal of the target in many times greater than the height of the explosion. It is also obvious that all nuclear explosions, data of which were used to receive the dependence of amounts of shattered glass from nominal overpressure of the shock wave, were made at low altitudes. So that the distance from them to the areas where is possible to consider the broken glass on the walls of entire buildings, was many more than the heights of these explosions (typical altitude of airburst for warhead with the energy of about 1 Mt TNT was approximately 1.5 km with a characteristic radius of glass break about 20 – 40 km, see [12, 14]). In such a case, into facets which are turned towards to the explosion, such as the walls of houses with windows, straight shock wave falls, plane of which is parallel to the plane of these facets. If the burst is high-altitude, as at Chelyabinsk, then right along the surface of the earth runs oblique shock wave (inclined to the surface), which in this scale is almost flat part of the spherical shock wave of explosion. And, the farther away from the epicenter, the greater becomes its slope.

The straight shock wave falls down on rooftops at the epicenter of the explosion, and to a first approximation, slides practically along their vertical walls interacting with them relatively weakly. At more detailed examination of this process should be taken into account a turn of the shock wave on the facets formed by a flat roof of house, or on a set of oblique edges formed by a peaked roof. This unfolded and oblique shock wave is reflected from the ground to form new oblique shock waves. And, in principle, the impact of these waves on walls of buildings and its windows can be determined by numerical simulations for a specific geometry and arrangement of buildings standing close. However, it's well known even without gasdynamics calculations that any oblique shock wave is a «weaker» than a straight, and its impact on the obstacle should be less significant. That is why the proportion of broken glass near the epicenter was far from absolute [6].

Such interaction, which occurs in the epicenter of explosion between a shock wave and a peaked roof of house, consisting of two or three facets, is implemented between wave and facets of standard multi-storey building with flat roof for average foreshortening of wave propagation, when height of explosion and distance to it are commensurate. Therefore, when the oblique shock wave runs along the earth, part of broken window panes will be significantly less than that of a straight shock wave from low-altitude explosion of the same energy.

So, oblique shock wave from high-altitude powerful explosion runs along the surface of the earth at a considerable distance from the epicenter, reflecting from the solid surface as another oblique shock wave. This type of reflection is called regular, and it can be represented schematically in the form of a letter V, where the slanting dashes depict the incident and reflected shock waves. At the time as the distance from the epicenter grows, the angle of the incident wave grows also, and at some point it leads to the inability to implement regular reflection. From this point it becomes so-called Mach reflection [17], which can be schematically represented as a letter Y – between the point (in the planar case) or line of intersection of oblique shock waves (in three dimensions) and a solid surface, so-called «Mach stem» arises (vertical dash of the letter Y), which is a straight shock wave. From that moment, a high-altitude airburst in its impact on the underlying surface becomes equivalent to a low-altitude explosion, and only then one can begin to compare part of broken glass from all previously existing sources with that, what happened after the explosion of Chelyabinsk meteoroid. And that is why the boundary conditions in [1, 2] were set at such a great distance, where were guaranteed the irregular or Mach reflection of a shock wave.

All of the above illustrates the simple and obvious fact that to determine the energy of high-altitude explosion through data of breakage of ground objects (including the part of broken glass), obtained at propagation of shock waves from low-altitude nuclear explosions, we should consider data from regions of Mach (irregular) reflection only. Glass shattered by waves can only be compared for comparable shock waves. Hence it becomes clear that where in source [5] was found «abnormal» level of overpressure in «caustic zone», which is not known for shock waves, there was realized the picture of the interaction of waves with obstacles, which alone can properly interpret the observed phenomena using available data of previously observed powerful explosions. For some desire this transition from regular to Mach reflection can be called «constructive interference» of the incident and reflected waves, resulting in a significant increase in real, not nominal pressure, as it wanted to do the authors of paper [5].

It remains now to consider only one question: why this «constructive interference» was seen in only one area of Chelyabinsk? To answer this question we turn to maps and satellite images of the terrain. Even on a large-scale map shown in Fig. 2 is seen that on line of propagation of the shock wave of Chelyabinsk explosion at a distance of about 35 km up to the Ice Palace «Urals Lightning» are almost entirely non built-up flat plains where are only fields or forests. This line pass then along the coastal part of Shershnevsky (Hornet) water basin, which was covered with ice during the explosion, and then line pass through Chelyabinsk city forest. If to move across this line, watching underlying areas with the aid of satellite images at full resolution, it could be seen that quantity of such obstacles as urban multi-storey buildings on this line can be counted up to Chelyabinsk Zinc Plant on the fingers of one hand.

Thus, the oblique shock wave from high-altitude explosion has propagated along the ground practically with no energy loss due to obstacles. And its high intensity was fixed through mass of broken glass only after transformation of the wave into a straight as waves of previously observed low-altitude explosions. To east, in the direction of the center of Chelyabinsk and its eastern regions on the way still an oblique shock wave has passed through large arrays of high multi-storey buildings, what was accompanied by the emergence of very large quantity of local incident and reflected waves and their interactions with each other and with new obstacles. Such process had to seriously affect the geometry of the shock wave near the ground, the average pressure levels and the picture of the destructions in those «shielded» areas of the city. And, apparently, the above-described model of transition on a smooth solid surface from regular to Mach reflection of shock wave had to transform into something much more complicated and chaotic.

It should also be noted that prior to the explosion over Chelyabinsk were only 2 cases of strong shock waves passing through a continuous urban development for a distance of several kilometers – in Hiroshima and Nagasaki, and in a much smaller scale, apparently, another 2 in Halifax and Texas City [18 – 20]. But the explosion energy in these cases was of a few thousand or tens of thousands times smaller than the explosion over Chelyabinsk, and path lengths of intense waves were at least by one or one and half order of magnitude smaller. Furthermore, the main part of buildings in these two Japanese cities was small one-storey or two-storey wooden houses [18]. This contrasts sharply with urban architecture in central and eastern districts of Chelyabinsk. Known correlations between the amount of broken glass and overpressure on shock wave were derived from data on the propagation of shock waves on the urban housing of Hiroshima and Nagasaki, or even on the bare steppe Semipalatinsk test site with detached buildings when the attenuation of these waves due to obstacles was weak. Therefore, these data are not fully adequate for all districts of Chelyabinsk, but only for direction to «Urals Lightning» and Zinc Plant, where there are vast areas with no high-rise buildings. This is true even without differences of impact of straight and oblique shock waves on obstacles.

In this context, it should also be noted that there is another factor from which the altitude of explosion has significantly affected to the impact on the underlying surface. This is a decrease in density of the atmosphere at the point of explosion by the growth of its height. For heights of 25 – 30 km, this factor can lead to a reduction in overpressure at the surface in several times. Influence of this factor is studied in detail in the article criticizing sources [11].

Further calculations showed that the condition p = 7.5 ± 0.5 kPa at a distance of 39.5 km from the explosion epicenter is equivalent to the condition p = 5.0 kPa at a distance of 80 ± 6 km, which is in good agreement with previous estimates of the model for the affected area. Thus, the previous analysis shows that the overpressure peak on the shock wave of 7 – 8 kPa in the region of Chelyabinsk in the neighborhood of the Zinc Plant [5] is the most accurate from available boundary conditions for solving the problem.

## IV. Calculation results of Chelyabinsk and Tunguska meteoroids' parameters and their explosion modes

Several tens of calculations of inputs into the atmosphere and explosions of Chelyabinsk and Tunguska meteoroids were made, and the results of 11 of them as the most representative are shown in Tables 1 – 4. Tables 1 – 2 show the influence of height changing in explosion point to the characteristics of Chelyabinsk meteoroid (ChM), as well as to the overpressure peak on the shock wave from the explosion at several distances from the epicenter. As stated in section II of this paper, the impact velocity of the object was 18.85 km/s. A minimum height of the explosion point, which was some approximation of the real Chelyabinsk height of blast, is, as shown above, 27.0 km. This value was increased in increments of 0.5 km in calculations with the boundary condition 5.00 kPa at distance of 80.0 km from the explosion epicenter. Nominal height of the atmosphere at this process assumed to be equal 90.0 km, which, according to [4], approximately corresponds to the beginning of its impact on the object of this type.

In Table 1 are shown: var – variant of calculation of Chelyabinsk meteoroid, H is height of explosion in kilometers, δ is entry angle in degrees, ρ is density of the object in kilograms per cubic meter, D is diameter of the object in meters, m is mass in megatons, $E_0$ is kinetic energy of the object entering the atmosphere in megatons of TNT, $E_e$ is explosion energy of the object in the same units.

**Table 1**

| var | H (km) | δ (º) | ρ (kg/m$^3$) | D (m) | m (Mt) | $E_0$ (Mt) | $E_e$ (Mt) |
|---|---|---|---|---|---|---|---|
| **ChM-1** | 27.0 | 7.22 | 870 | 158.5 | 1.81 | 76.6 | 52.6 |
| **ChM-2** | 27.5 | 7.19 | 740 | 167.5 | 1.82 | 77.3 | 54.4 |
| **ChM-3** | 28.0 | 7.17 | 630 | 177 | 1.83 | 77.8 | 56.2 |
| **ChM-4** | 28.5 | 7.14 | 540 | 187 | 1.85 | 78.3 | 57.8 |
| **ChM-5** | 29.0 | 7.11 | 460 | 197.5 | 1.86 | 78.9 | 59.6 |

As shown in Table 1, increasing the height of the air blast from 27 to 29 km under these conditions results in a decrease in the angle of entry into the atmosphere at 0.1 º and drop in the density of the object in 1.9 times – from 870 to 460 kg/m$^3$. The diameter of it is growing at 25% – about from 160 to 200 m, while increasing its mass and kinetic energy at the input into the atmosphere by 3 %: m = 1.81 – 1.86 Mt , $E_0$ = 76.6 – 78.9 Mt of TNT. Thus, this energy is increased by 2.3 Mt, while the explosion energy $E_e$ increases more rapidly – from 52.6 to 59.6 Mt, that is by 7.0 Mt or 13 % of initial value. This is due to the reduction of energy losses less dense meteoroid during braking at higher altitudes that is at a lower density of the atmosphere.

Table 2 shows the values of the main factor of the shock wave from explosion on the ground obstacles at these distances – the overpressure peak. Here: var – variant, p is overpressure peak on the shock wave in kilopascals at a distance L from the explosion, measured in kilometers along the ground and demonstrated in the column to the left of the pressure.

**Table 2**

| var | $L_0$ (km) | $p_0$ (kPa) | $L_1$ (km) | $p_1$ (kPa) | $L_2$ (km) | $p_2$ (kPa) | $L_3$ (km) | $p_3$ (kPa) | $L_4$ (km) | $p_4$ (kPa) |
|---|---|---|---|---|---|---|---|---|---|---|
| **ChM-1** | 0 | 11.6 | 20 | 9.4 | 35 | 8.1 | 39.5 | 7.67 | 80 | 5.0 |
| **ChM-2** | 0 | 11.4 | 20 | 9.3 | 35 | 8.0 | 39.5 | 7.60 | 80 | 5.0 |
| **ChM-3** | 0 | 11.2 | 20 | 9.2 | 35 | 7.9 | 39.5 | 7.53 | 80 | 5.0 |
| **ChM-4** | 0 | 11.0 | 20 | 9.0 | 35 | 7.8 | 39.5 | 7.46 | 80 | 5.0 |
| **ChM-5** | 0 | 10.8 | 20 | 8.9 | 35 | 7.7 | 39.5 | 7.39 | 80 | 5.0 |

Maximum overpressure peak is attained on the shock wave at the epicenter (at $L_0$ = 0). The more powerful and higher is the explosion, the lower is this overpressure. The distance $L_1$ = 20 km is characteristic for Tunguska explosion, which is compared with Chelyabinsk blast, the distance $L_2$ = 35 km roughly corresponds to the length between its epicenter and the center of Chelyabinsk (as well as the Ice Palace «Urals Lightning»). At the distance of $L_3$ = 39.5 km is located Chelyabinsk Zinc Plant, the distance $L_4$ = 80 km is the length to the border of a circular area with overpressure of 5.0 kPa, which approximates the real zone of destruction. Nominal overpressures on a straight shock wave have not reached 12 kPa even in the epicenter, at a distance of 35 km (in the center of Chelyabinsk) it was about 8 kPa and below, and in the area of Zinc Plant the pressure level is near 7.7 – 7.4 kPa. As stated earlier, these magnitudes may be reached here after the realization of Mach reflection and arising of straight shock wave. These conditions are comparable to those that have realized at low-altitude explosions. So how exactly these conditions for straight shock wave are employed in all the data on broken windows, it is this area, which should be used to fine tune the parameters of the computational model that describes in the best way what happened in reality.

It follows from Table 2 that the overpressure on the wave of 7.50 kPa at the distance of 39.5 km (average value for this region according to the source [5]) is realized when the height of the explosion is 28.2 km. This is perfectly consistent with the preliminary estimates: «the interval from 27 to 30 km, but closer to the lower boundary of heights», as well as with the data source [9]. Then the meteoroid explosions were calculated at the altitude of 28.2 km and at the overpressure peak on the shock wave from 7.0 to 8.0 kPa at the distance of 39.5 km. At the same time was also adjusted the nominal height of the atmosphere – it was increased from 90.0 to 91.2 km. Iterative computational procedure showed that this height of the atmosphere corresponds to the beginning of its impact for the base version of Chelyabinsk meteoroid ChM-7 (see Table 3), which creates an explosion overpressure of 7.50 kPa at a specified distance. Further, all calculations were carried out precisely at such nominal height of atmosphere.

Calculated data for three variants of the Chelyabinsk meteoroid at the overpressures on the wave of 7.00 kPa, 7.50 kPa and 8.00 kPa at the distance $L_3$ = 39.5 km are shown in the first three lines of Tables 3 and 4. The values and the designations are the same as before. Bold fonts are used for basic variants of Chelyabinsk (ChM-7) and Tunguska (TM-1) meteoroids.

**Table 3**

| var | v (km/s) | i (º) | H (km) | δ (º) | ρ (kg/m$^3$) | D (m) | m (Mt) | $E_0$ (Mt) | $E_e$ (Mt) |
|---|---|---|---|---|---|---|---|---|---|
| **ChM-6** | 18.85 | 5.00 | 28.2 | 7.22 | 635 | 173 | 1.71 | 72.7 | 52.1 |
| **ChM-7** | **18.85** | **5.00** | **28.2** | **7.22** | **570** | **182.5** | **1.82** | **77.4** | **56.8** |
| **ChM-8** | 18.85 | 5.00 | 28.2 | 7.22 | 515 | 193 | 1.93 | 82.1 | 61.8 |
| **TM-1** | **18.72** | **– 5.00** | **8.25** | **50.5** | **570** | **105** | **0.35** | **14.6** | **14.4** |
| **ChM-9** | 19.00 | 5.00 | 28.2 | 7.25 | 580 | 181 | 1.80 | 77.5 | 56.7 |
| **TM-2** | 18.87 | – 5.00 | 8.33 | 50.0 | 580 | 104.5 | 0.35 | 14.8 | 14.6 |

Entry angles δ in all three variants of calculations (ChM-6 – ChM 8) coincide with each other up to the third digit after the decimal point. The density of the object ρ at overpressure increasing at a predetermined range is decreased by 23 % – from 635 to 515 kg/m$^3$, and the diameter D is increased by 11.5 % from 173 to 193 m. Accordingly, mass m is increased substantially by the same amount, and the explosion energy $E_e$, providing the required overpressure, is increased by 19 %. Thus, a dynamic equilibrium is reached at the point of the object destruction: greater but less dense object is destroyed at the same height as the smaller but denser body. Growth of the masses and energies of the object under increasing the overpressure on the shock wave at a predetermined distance from the explosion is quite natural process. At the overpressure of 7.50 kPa the calculated mass of base variant of Chelyabinsk meteoroid ChM-7 is 1.82 Mt at a diameter of 182.5 m, and the energy of its explosion $E_e$ is 56.8 Mt of TNT, which is only 1.0 Mt (1.8 %) less than in early calculations [1, 2]. Its total energy $E_0$ was 2.5 % greater due to higher initial flight speed.

**Table 4**

| var | $L_0$ (km) | $p_0$ (kPa) | $L_1$ (km) | $p_1$ (kPa) | $L_2$ (km) | $p_2$ (kPa) | $L_3$ (km) | $p_3$ (kPa) | $L_4$ (km) | $p_4$ (kPa) |
|---|---|---|---|---|---|---|---|---|---|---|
| **ChM-6** | 0 | 10.3 | 20 | 8.5 | 35 | 7.3 | 39.5 | 7.00 | 73.9 | 5.0 |
| **ChM-7** | **0** | **11.1** | **20** | **9.1** | **35** | **7.9** | **39.5** | **7.50** | **80.0** | **5.0** |
| **ChM-8** | 0 | 12.0 | 20 | 9.8 | 35 | 8.4 | 39.5 | 8.00 | 85.5 | 5.0 |
| **TM-1** | **0** | **82.9** | **20** | **30.0** | **35** | **11.8** | **39.5** | **9.8** | **63.4** | **5.0** |
| **ChM-9** | 0 | 11.1 | 20 | 9.1 | 35 | 7.8 | 39.5 | 7.50 | 80.0 | 5.0 |
| **TM-2** | 0 | 81.6 | 20 | 30.0 | 35 | 12.0 | 39.5 | 9.9 | 63.8 | 5.0 |

We turn now to Tunguska meteoroid and to its explosion. Earlier in paper [10] it was proved the generality of the origin of Tunguska and Chelyabinsk meteoroids, as they were members of the same family of cometary debris [21]. This leads to proximity of orbits of these two objects. Therefore, to evaluate parameters of Tunguska meteoroid there was used the same orbit as for Chelyabinsk meteoroid with the sole exception – angle of inclination of the orbit plane has got the opposite sign (see Table 3). The estimations showed that, while maintaining an argument perihelion, this orbit's modification can approximately provide its intersection with the Earth orbit in late June – early July that is in the first window of rapprochement with cluster of close orbits of Tungus family [10, 21].

The explosion of Tunguska meteoroid (TM in Tables 3 and 4) occurred June, 30 1908, in the first window of approach with another position of the Earth axis to ecliptic plane and with such velocity vector position of the object, which leads to a mirror image of it relative to the velocity vector of the planet compared to what it was in February 2013. The Tunguska explosion was considerably northerly of Chelyabinsk – its coordinates were: 60.89° north latitude and 101.90° east longitude [1, 2]. Local time of the explosion was 7:14:30, solar time – 7:02:06. All of these factors combine to affect the increase of entry angle of Tunguska meteoroid, which at geodetic azimuth of 279° (9° inclination of the trajectory parallel to the north) [1, 2] was equal to 50.5°.

Such short and steep tracks are computed much easier and faster than long and flat trajectories. At the same time for more accurate data, in contrast to earlier calculations [1, 2], there was applied the calculation module at atmospheric part of the trajectory. On such trajectory this module has very small influence on the final result, as the effect of the atmosphere on the path up to the explosion of the object is minimal. This is evident from the fact that its energy has decreased on the atmospheric part of the trajectory by only 1.4 % (see line TM-1 in Table 3). Interactive module, which describes the destruction of meteoroids in the atmosphere [3, 4], operates in the framework of «flat» Earth. That also introduces additional errors in the results of calculations of long and flat trajectories. Therefore, the calculated entry angle of Chelyabinsk meteoroid is different from the real, and it can be regarded only as estimation.

These calculations of short and steep trajectory of Tunguska meteoroid are practically free from these errors. And all possible errors are determined only by the deviation of its computed orbit from real, that, of course, could and should be somewhat different from Chelyabinsk meteoroid's orbit not only by inclination angle with the ecliptic plane, but these deviations should be sufficiently small. However, it should be assumed that to obtain quantified results in calculating the characteristics of the Tunguska meteoroid we may use this orbit. Because of this, as well as of unity of meteoroids origin, the average density (about 570 kg/m$^3$, see lines ChM-7 and TM-1 in Table 3) were the same. Therefore, the value of density was enough correctly computed and for Chelyabinsk meteoroid.

For anyone representing the processes of comet's nuclei and their debris evolution should be clear that known from other sources the value of its density – 3300 kg/m$^3$ (see, for example, [6]) is the density of external crust of meteoroid. This crust is formed due to solar ablation of snow-ice composite which is contaminated by chondrite [22]. Therefore, the density of a relatively thin crust does not characterize the average density of the object before destruction. The thickness of this crust should be of order of meter that follows from parameters of the largest surviving fragment of Chelyabinsk meteoroid, see, for example, its photo [23]. It is obvious that only a small portion of this crust may survive the explosion, while snow and ice – main part of the meteoroid material should to evaporate. And we should not to judge about the average density of large object watching only these insignificant residuals of this thin surface layer.

In general, it follows from the calculations that by increase (compared with earlier calculations [1, 2]) of the entry velocity of Tunguska meteoroid up to 18.72 km/s (due to the growth of the refined speed of Chelyabinsk meteoroid up to 18.85 km/s) the main parameters of the first object and characteristics of its explosion changed very slightly. Only height of the explosion has increased by 7 % up to 8.25 km, and the density – on 14% to 570 kg/m$^3$ (see [1, 2]). Explosion energy was increased only by 0.7 %, which, of course, does not go beyond the errors of the method. Estimated average density of comet debris, though increased slightly, but remained at the same level, consistent with the data for comets and Earth snows [1, 2].

The diameter of Tunguska meteoroid was approximately in 1.75 times smaller than of Chelyabinsk object, its mass was in 5.2 times less, and the energy of the explosion – in 3.9 times less (see lines ChM-7 and TM-1 in Table 3). But, since its explosion occurred at the height of 3.4 times lower, the impact on the underlying surface was not an example of stronger. The overpressure peak in the epicenter on a straight shock wave is estimated in 7.5 times more than at the explosion of Chelyabinsk meteoroid (see lines ChM-7 and TM-1 in Table 4). As reported, the roofs were not damaged in the vicinity of the epicenter of Chelyabinsk explosion, when the overpressure maximum at the earth surface was of about 10 – 11 kPa. Overpressure on the glass was significantly lower what was explained in detail in the previous section of this paper. And because of this there was relatively few of broken glass. In contrast, in taiga near Stony Tunguska River was the region of full tree-felling at the radius not less than 20 km except the epicenter, where there was a dead forest from tree trunks completely without branches [24]. Boundary of the overpressures equality lies from the epicenter at the distance of about 51.5 km away. At greater distances the stronger wave was in a much more powerful and much more high-altitude Chelyabinsk explosion.

From these computations is followed that the level of overpressure peak on the shock wave of Tunguska explosion 5.0 kPa is reached at a distance from the epicenter at 63.5 – 64 km (see lines TM-1 and TM-2 in Table 4). And exactly at this distance there is the closest village – Vanavara. Eyewitnesses – Vanavara residents have reported the following: «Then it turned out that many of the windows were broken» [24]. This is consistent with the calculated data shown in Tables 3 and 4. It should also be noted that the estimates of the energy of the Tunguska explosion from seismograms leads to the value of its energy 12.5 ± 2.5 Mt, and from barograms – 12 ± 2.5 Mt [25, 26], that is in good agreement with that obtained in this calculation of its magnitude, which was equal to 14.4 Mt (see line TM-1 in Table 3).

Increase of entry speed of Chelyabinsk meteoroid from 18.85 km/s to 19.00 km/s had in general to a negligible effect on its characteristics (see lines ChM-7 and ChM-9 in Table 3). Slightly reduced size, slightly increased density and angle of entry, and the two concerned energy values have not changed. Overpressure peaks on the shock wave also remained virtually unchanged (see lines ChM-7 and ChM-9 in Table 4). Approximately similar but somewhat more significant is the influence on the characteristics of Tunguska meteoroid in the case of corresponding increase in speed with 18.72 to 18.87 km/s (see lines TM-1 and TM-2 in Tables 3 and 4). Its explosion energy is raised by 0.2 Mt to 14.6 Mt, which is greater by 1.4% than at lower speed.

## V. Discussion of results

Thus, in the morning February 15, 2013 some celestial body has exploded in the sky over Chelyabinsk at a height near 28 km (28.2 km for modeling point of blast). Its size was of approximately 180 – 185 m, density was of about 570 kg/m$^3$ and mass – of about 1.8 Mt. Energy of the explosion was 56.8 ± 4.9 megatons of TNT with accounting of maximum overpressure error on shock wave ± 0.5 kPa in the region of Zink Plant. Hence, the explosion energy in the sky at Chelyabinsk was almost equal to the energy of the most powerful thermonuclear explosion of so-called Tsar bomb, which amounted to 58 Mt (other designations – AN602, Kuzka's mother), produced by the Soviet Union October 30, 1961 at Novaya Zemlya [27]. Determination of Chelyabinsk explosion energy by acoustic methods leads to a value fully coinciding with this magnitude calculated there with lower error – 56.8 ± 1.1 Mt [28]. This seems to be the most accurate estimate of the energy of this explosion.

Over 104.5 years before this, June 30, 1908 some meteoroid has exploded on the Stony Tunguska River, which was much smaller, however, it is still considered as the largest celestial body that entered the Earth atmosphere in historic times. This celestial body had the same density, but its size was 105 m, and mass – 0.35 Mt. The energy of explosion was 14.4 Mt, but because of that the height at which this explosion has occurred was in 3.4 times lesser, that is 8.25 km, the impact on the underlying surface at that time was not an example of a stronger. The calculated data of Tunguska incident are in good agreement with those previously obtained by several generations of researchers for decades of work on this problem: the energy of the explosion from 7 to 17 Mt at the altitude of between 6.5 and 10.5 km [29]. The calculated explosion energy is also within the boundaries of 10 – 15 Mt defined by seismic data and barograms [25, 26].

That these objects were members of the same family of cometary debris was stated as a conjecture in papers [1, 2, 21] and was proved in article [10]. Owing to the proximity of orbits, which have the members of this Tungus family [21], the characteristics of any from these objects may be evaluated by the method of this work with high degree of accuracy at the minimum information about them. The same can be done with the characteristics of objects the other related family, which was named as Eagle (L'Aigle) family [21]. Thus, this work is leading us back to the origins of a series of articles [1, 2, 10, 14, 21, 22, 28] and makes more exact their quantitative basis, and confirms again the concept that there are two families of cometary debris which were threatened to the world in the historical past and still are threatening up to the present time [10, 30].

However, there are very great differences between the results presented here, as well as in the papers [1, 2, 10, 14, 21, 22, 28] and the data described in two large review articles on Chelyabinsk incident [5, 6], where the preliminary results of studies of large scientific groups are presented. These differences are not only quantitative and qualitative, but even ideological. Because of this a separate paper was devoted to analyze and critique of the views of opponents [11].

## Conclusions

1. The results of calculations by the mathematical model that relates the parameters of celestial bodies motion in the spheres of activity of the Sun and the Earth, with the mass-energy characteristics of these objects and their explosion modes during the destruction in the atmosphere, turned well matched with the data obtained from observations.
2. Calculations have shown that size of Chelyabinsk meteoroid was equal to 182.5 meters, and its mass was close to 1.82 megatons. Energy of explosion was 56.8 ± 1.1 megatons of TNT.
3. Size of Tunguska meteoroid was close to 105 m, its mass was 0.35 megatons, while the energy of the explosion was about 14.5 megatons of TNT.
4. Due to the common origin of these two celestial bodies their average density was equal – about of 570 $kg/m^3$.
5. This mathematical model may also be used for calculating the characteristics of other celestial bodies entering the Earth atmosphere to replace guesses, assumptions and myths by scientific data.